\documentclass{ws-procs961x669}            
\usepackage{graphicx}
\usepackage{mathrsfs}

\def\beq{\begin{equation}}
\def\eeq{\end{equation}}

\newcommand{\bo}{\raise-1mm\hbox{\Large$\Box$}}

\newcommand{\f}[2]{\frac{#1}{#2}}

\newcommand{\la}{\langle}
\newcommand{\ra}{\rangle}
\newcommand{\w}{\omega}
\newcommand{\kp}{\kappa}
\newcommand{\be}{\begin{equation}}
\newcommand{\ee}{\end{equation}}
\newcommand{\bea}{\begin{eqnarray}}
\newcommand{\eea}{\end{eqnarray}}

\begin{document}

\title{Reflections on a Black Mirror}
\author{Michael R.R. Good}

\address{Physics Department, Nazarbayev University,\\
Astana, Republic of Kazakhstan\\
E-mail: michael.good@nu.edu.kz}



\begin{abstract}
A black mirror is an accelerated boundary that produces particles in an exact correspondence to an evaporating black hole.  We investigate the spectral dynamics of the particle creation during the formation process.
\end{abstract}


\bodymatter


\vspace{0.5cm}
To understand particle creation from black holes, it is essential to consider the time-dependent formation phase. Recently, it was shown\cite{paper1, paper2,Good:2016oey} that there is an exact correspondence between a black hole and a moving mirror.  That is, for a massless minimally coupled scalar field in a (1+1)D spacetime in which a black hole forms from the collapse of a null shell, and for the quantum amplification of modes for this field which results from an accelerating moving mirror with a particular trajectory in (1+1)D, the particle creation is the same. The moving mirror in this case is called a \textit{black mirror}. 

First, before we investigate the black mirror's early-time particle creation, allow us to take a moment to inquire about a system   which emits thermal radiation forever\cite{Carlitz:1986nh}.  In this case there is no time dependence for the particle creation.  This system, unlike the black mirror, is in equilibrium.  The particles residing in each frequency mode are Planckian:
\be   N_\omega = \frac{1}{e^{2\pi \omega/\kappa}-1}, \label{Planckian} \ee
with temperature $T = \kappa / 2\pi$. By construction, the immutable constant emission of energy ($F = \kappa^2/48\pi$) of this Carlitz-Willey \cite{Carlitz:1986nh} moving mirror is echoed by the eternal constant emission of Planckian distributed particles.  The relevant beta Bogoliubov coefficient for this eternally thermal moving mirror (henceforth called the \textit{equilibrium mirror}) is
\be \beta_{\omega\omega'} = -\frac{1}{2\pi\kappa}\sqrt{\frac{\omega}{\omega'}}e^{-\pi \omega/2\kappa}\left(\frac{\omega'}{\kappa}\right)^{-i\omega/\kappa} \Gamma\left[\frac{i\omega}{\kappa}\right]. \label{betaCW} \ee
One can see the form of the Planckian distribution of Eq.~\ref{Planckian} by complex conjugating:
\be |\beta_{\w\w'}|^2 = \f{1}{2\pi\kp\w'} \f{1}{e^{2\pi \w/\kp} - 1} \;. \label{betasquared} \ee
Albeit, there is a pre-factor of $1/2\pi\kp\w'$ which is responsible for the total infinite production of particles, residing in mode $\w$, when one integrates over $\w'$:
\be  \langle N_\omega \rangle = \int_0^\infty d \omega' |\beta_{\w \w'}|^2  \;.  \label{totalparticles} \ee
To avoid infinite particle production, and to remove this pre-factor in Eq.~\ref{betasquared}, one most localize the production of particles in time and frequency.  Following Hawking\cite{Hawking:1974sw}, one uses orthonormal and complete wave packets to packetize and localize the beta Bogoliubov coefficient in Eq.~\ref{betaCW}.  Then, one has
packetized beta Bogoliubov coefficients \cite{Fabbri:2005mw, Good:2015nja, Good:2013lca},
\be
\beta_{jn\w'} =
\f{1}{\sqrt{\epsilon}}\int_{j\epsilon}^{(j+1)\epsilon}
d\w \; e^{\frac{2\pi i \w n}{\epsilon}} \beta_{\w\w'} \;.
\label{beta-packet}
\ee
The values of $n$ and $j$ designate the time and frequency bins respectfully.  The total number of particles in any particular packet is then,
\bea
\la N_{jn}\ra &=& \int_0^\infty d\w' |\beta_{jn,\w'}|^2   \;,
\nonumber
\\
&=& \int_0^\infty d\w' \int_{j \epsilon}^{(j+1)\epsilon}
\frac{d \w_1}{\sqrt{\epsilon}} \int_{j \epsilon}^{(j+1)\epsilon}
\frac{d \w_2}{\sqrt{\epsilon}} e^{2 \pi i(\w_1- \w_2)n/\epsilon}
\beta_{\w_1 \w'} \beta^{*}_{\w_2 \, \w'}  \;,
\nonumber
\\
&=& \frac{1}{\epsilon}\int_{j \epsilon}^{(j+1)\epsilon}
d \w_1 \int_{j \epsilon}^{(j+1)\epsilon}
d \w_2 e^{2 \pi i(\w_1- \w_2)n/\epsilon}
\langle N_{\w_1\w_2} \rangle  \;,
\label{Njn}
\eea
where
\be \langle N_{\w_1\w_2} \rangle  \equiv \int_0^\infty d \w' \beta_{\w_1\w'}\beta^*_{\w_2 \w'}\;. \ee
The particles are localized in the range of frequencies $j \epsilon \leqslant \omega \leqslant (j+1) \epsilon$.  They peak in the range of times $ (2\pi n - \pi)/\epsilon \leqslant u \leqslant (2 \pi n + \pi)/\epsilon$.  Further details about these wave packets are discussed in Fabbri\cite{Fabbri:2005mw} and Good-Anderson-Evans\cite{Good:2013lca}. In the case of the equilibrium mirror, the packetized particle count is found\footnote{This delta function comes about with help from using the integral: $\int_0^\infty d\w' \w'^{i(\w_2-\w_1)-1} = 2\pi \delta(\w_2-\w_1)$.} exactly:
\be \langle N_{\w_1\w_2} \rangle = \frac{1}{e^{\w_2/T}-1}\delta(\w_2-\w_1). \label{Nww}\ee
To see how there is no time dependence, one integrates with the Dirac delta function by plugging Eq.~\ref{Nww} in the last line of Eq.~\ref{Njn}. This causes the time dependence to cancel out because the $n$ time bin values, in the term $e^{2 \pi i(\w_1- \w_2)n/\epsilon}$, are negated by the Dirac delta function. Therefore, the particle count, for the equilibrium mirror, becomes eternally time independent, $\langle N_{j n} \rangle \rightarrow \langle N_j \rangle$: 
\be \langle N_{j} \rangle = \frac{1}{\epsilon} \int_{j\epsilon}^{(j+1)\epsilon} d \w_1 \frac{1}{e^{\w_1/T} - 1}. \ee
This integral is easy to do and the final result is:
\be \langle N_{j} \rangle = \frac{1}{2\overline{\epsilon}} \log \left[\sinh \left(j+1) \overline{\epsilon}\right) \text{csch}\left(j \overline{\epsilon}\right)\right]-\frac{1}{2} \ee
where $\overline{\epsilon} \equiv \epsilon/2T$, or alternatively expressed, 
\be \la N_{j} \ra =  \frac{T}{\epsilon} \, \log \left[ \frac{e^{(j+1) \epsilon/T} - 1}{e^{j \epsilon/T} - 1} \right] - 1  \;.  \label{Njn-thermal} \ee
This is the exact Planckian distribution. It is this expression, Eq.~\ref{Njn-thermal}, which will allow us to numerically confirm, the late-time thermal emission of the black mirror.  This is the reason for investigating the equilibrium mirror first.  The equilibrium emission will help us explore non-equilibrium emission.

It is easy to check, that in an expansion of $\epsilon$, one can obtain the more familiar Planckian form.  Keeping the center frequency, $\omega_j \equiv (j+1/2)\epsilon$, constant, and doing the series expansion, then $\langle N_{j} \rangle \rightarrow \langle N_{\w_j} \rangle$, and Eq.~\ref{Njn-thermal} becomes the usual Planckian distribution: 
\be \la N_{j} \ra = \la N_{\omega_j} \ra + O(\epsilon^2), \ee
where 
\be \la N_{\omega_j} \ra \equiv \frac{1}{e^{\omega_j/T} - 1} \label{NjnPlanckian}\ee
and $T = \kappa/2\pi$. Eq.~\ref{NjnPlanckian} is the discrete counterpart of Eq.~\ref{Planckian} and for it to be accurate, one requires a small $\epsilon < 1$ wave packet parametrization.  The exact Planckian distribution, Eq.~\ref{Njn-thermal}, will be more useful than Eq.~\ref{NjnPlanckian}, as an anchor for investigating the time evolution of non-thermal systems, which eventually evolve into thermal equilibrium.   

Do we have an intuition about the exact Planckian distribution, Eq.~\ref{Njn-thermal}? What choice to make for $\epsilon$ as an intuitive parametrization for the wave packets?  Our preference is that it should be small, where Eq.~\ref{NjnPlanckian} holds, yet not so small, that time resolution is too coarse. Fine-grained frequency \textit{and} time resolution is required to analyze the spectral dynamics of the formation to a black hole.  To this end, we ask what is the typical angular frequency, $\epsilon_T$ for a Hawking quanta in thermal equilibrium? Yan\cite{Yan2000} asserts that the general thermal de Broglie wavelength is: 
\be \lambda_T = \frac{2 \sqrt{\pi}}{T} \left(\frac{\Gamma(d/2 +1)}{\Gamma(d+1)}\right)^{1/d}, \label{yen}\ee
using a massless linear dispersion relation with space dimensions $d$, and $\hbar=c=k_B=1$. Knowing this value, we could pick a packetization $\epsilon$ that is smaller, $\epsilon < \epsilon_T$, in order to resolve well in frequency, yet not be so small that we can't also have reasonably fine-grained time resolution for a non-eternal process. In our $1+1$ dimensional setting, Eq.~\ref{yen} gives,
\be \lambda_T = \frac{\pi}{T}, \quad \text{and} \quad f_T \lambda_T = 1 \rightarrow \frac{\epsilon_T}{2\pi} \frac{\pi}{T} = 1 \quad \rightarrow \quad \epsilon_T = 2 T. \label{thermalfreq}\ee 
We are then justified in picking $\epsilon < \epsilon_T = 2 T$, as a wave packet parametrization for $\epsilon$ for good frequency resolution.  Note that $\overline{\epsilon} \equiv \epsilon/2T = \epsilon/\epsilon_T$. Let us pick
\be \epsilon_{gr} = T \ln \frac{1+\sqrt{5}}{2} = T \ln \phi, \label{golden} \ee
where $\phi = 1.618$ is the Golden Ratio, such that $ T \ln \phi \approx T/2$.  This means that our choice of $\epsilon_{gr}$ is about $4$ times smaller than the general thermal de Broglie angular frequency, Eq.~\ref{thermalfreq}, $\epsilon_T = 2 T$.  It also means a 5\% relative error between Eq.~\ref{NjnPlanckian} and Eq.~\ref{Njn-thermal}. That is, 5\% error between the approximate familiar Planckian form and the exact Planckian form, when $j=1$.

   However, the actual intuitive usefulness of our choice of $\epsilon = \epsilon_{gr}$, is that it gives on average, \textit{one particle per packet}, in the lowest non-diverging $j=1$ frequency bin, for $\la N_{j} \ra$:
\be \la N_{j=1} \ra = 1. \ee
This is checked by solving Eq.~\ref{Njn-thermal}, using $j=1$ and $\epsilon = \epsilon_{gr}$.  This also happens to be the same assumption as supposing that the total amount of particles detected on average is exactly $2$ particles, that is:
\be\sum_{j=2}^\infty\la N_{j} \ra = 1, \ee
or, of course,
\be \sum_{j=1}^\infty\la N_{j} \ra = 2. \ee

\begin{figure}[ht]
\begin{minipage}[b]{0.45\linewidth}
\centering
 \rotatebox{90}{\includegraphics[width=\textwidth]{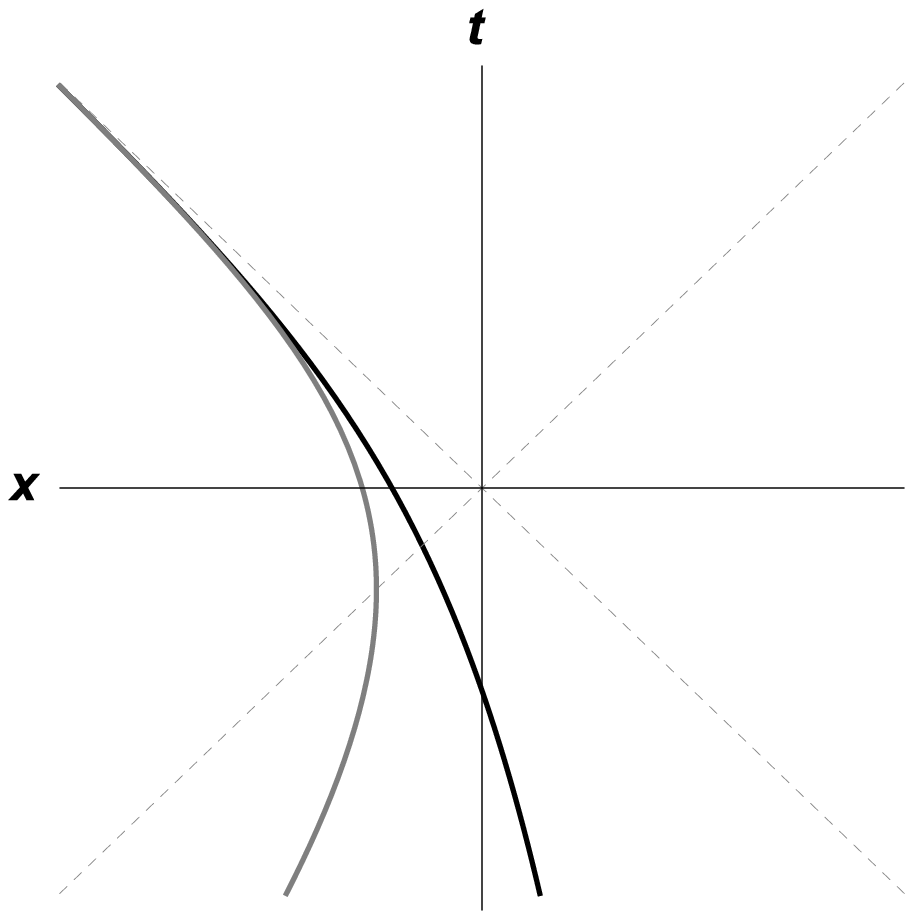}}
\caption{\label{fig:omexST} The black mirror in a spacetime diagram (the black trajectory) is overlaid with the equilibrium mirror (the gray trajectory).  The accelerations are time dependent.  The horizon is at $t=-x$.  We have set $\kp=1$.}
\end{minipage}
\hspace{0.5cm}
\begin{minipage}[b]{0.45\linewidth}
\centering
\includegraphics[width=\textwidth]{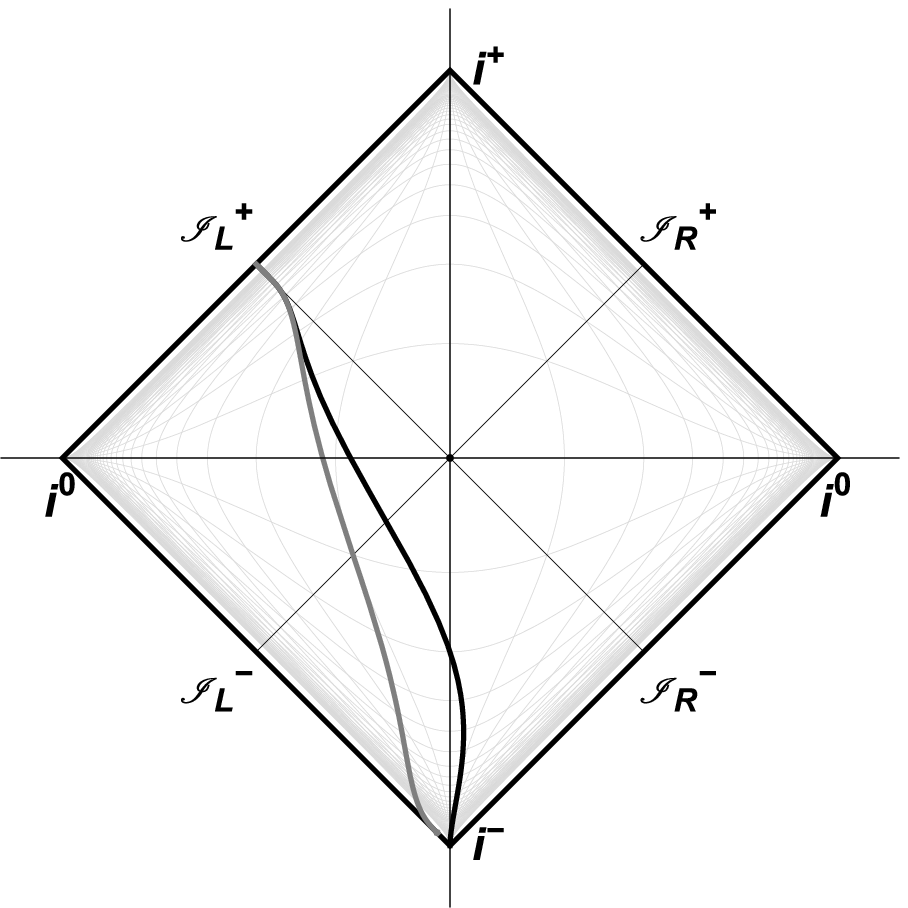}
\caption{\label{fig:omexPR} The two mirrors in a Penrose diagram.  Notice that both start asymptotically inertial. The black mirror starts at rest, while the equilibrium mirror starts at the speed of light. $\kp = 1$.   }
\end{minipage}
\end{figure}

In this note, up until now, we have been looking at the equilibrium mirror.  We developed an intuitive tool set to understand the thermal equilibrium where particle emission is discrete and one particle per packet is observed.  Now we are in a better position to time resolve the early-time, non-thermal history of the black mirror.    The black mirror trajectory is given by\cite{paper1, paper2} 
\be z(t) =  -t - \f{W(2 e^{-2 \kappa t})}{2\kappa},  \label{trajectory} \ee
 where $W$ is the Lambert W function, and the horizon is taken to be on the $t=-x$ null ray.
It is plotted in Fig. (\ref{fig:omexST}) and it begins at rest, at $z = +\infty$ and at late times asymptotically approaches the speed of light.
At late times this trajectory is very similar to the equilibrium mirror\footnote{See Good-Anderson-Evans\cite{Good:2013lca} or Good\cite{Good:2012cp} for the explicit equilibrium $z(t)$ trajectory.}.  The black mirror has particles in a non-thermal distribution at early times and they only become Planckian at late times.

The black mirror has beta Bogoliubov coefficients, 
      \be      \beta_{\w \w'} =  \frac{ \sqrt{\w \w'}}{2 \pi \kappa (\w +\w')} \, e^{- \pi \w/2\kappa} \left( \frac{\w + \w'}{\kp} \right)^{-i \w/\kappa} \Gamma\left[\frac{i \w }{\kp}\right],  \label{beta} \ee
with different form than equilibrium mirror beta coefficients of Eq.~\ref{betaCW}.
The total number of particles produced with frequency $\w$ is Eq.~\ref{totalparticles} and therefore complex conjugating, one finds
\be |\beta_{\w\w'}|^2 = \f{\w'}{2\pi\kp(\w+\w')^2} \f{1}{e^{2\pi \w/\kp} - 1}. \label{bmb2} \ee
Like the equilibrium mirror, we also see the explicit Planckian signature. It's easy to see that in the late time limit where $\w'\gg \w$, then Eq.~\ref{bmb2} goes to Eq.~\ref{betasquared}, 
\be |\beta_{\w\w'}|^2 \sim \f{1}{2 \pi\kp\w'} \f{1}{e^{2\pi \w/\kp} - 1} \;,
\ee
and there is a thermal distribution of particles with temperature $T=\kp/2\pi$.  Late time thermality was found $40$ years ago by Davies-Fulling\cite{Davies:1976hi} \cite{Davies:1977yv} for particular late-time behaved mirrors.  

\begin{figure}[ht]
\begin{minipage}[b]{0.45\linewidth}
\centering
\includegraphics[width=\textwidth]{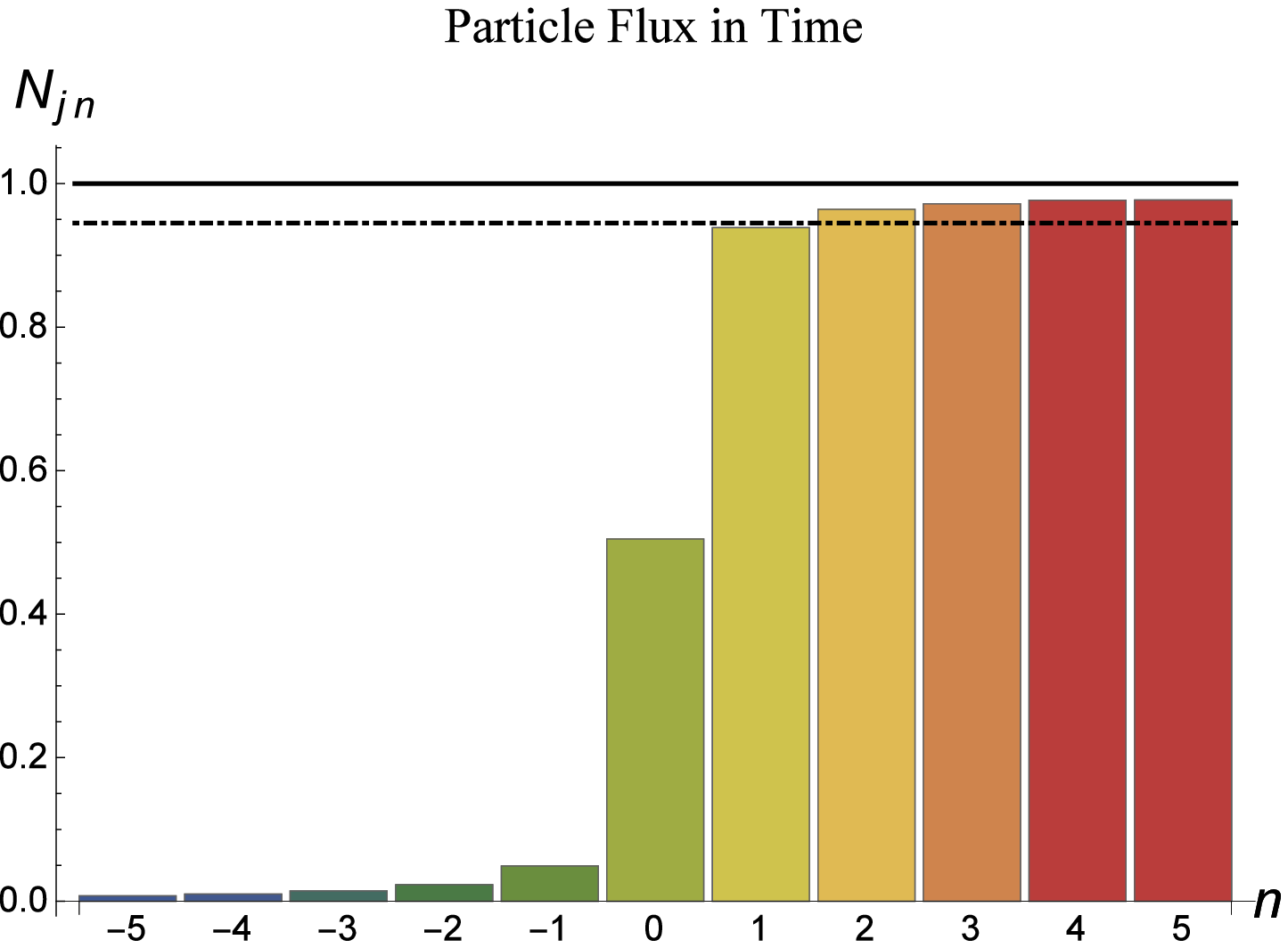}
\caption{\label{fig:pfTIME}Black mirror particle flux in time. The solid black line at the top is
the number of particles produced in an exact Planckian distribution, Eq.~\ref{Njn-thermal}, $\la N_{j=1} \ra = 1$. The dotted-dashed line is the Planckian distribution, Eq.~\ref{NjnPlanckian}. Here $j=1$, $\epsilon = T \ln \phi$, $\kp = 1$. }
\end{minipage}
\hspace{0.5cm}
\begin{minipage}[b]{0.45\linewidth}
\centering
\includegraphics[width=\textwidth]{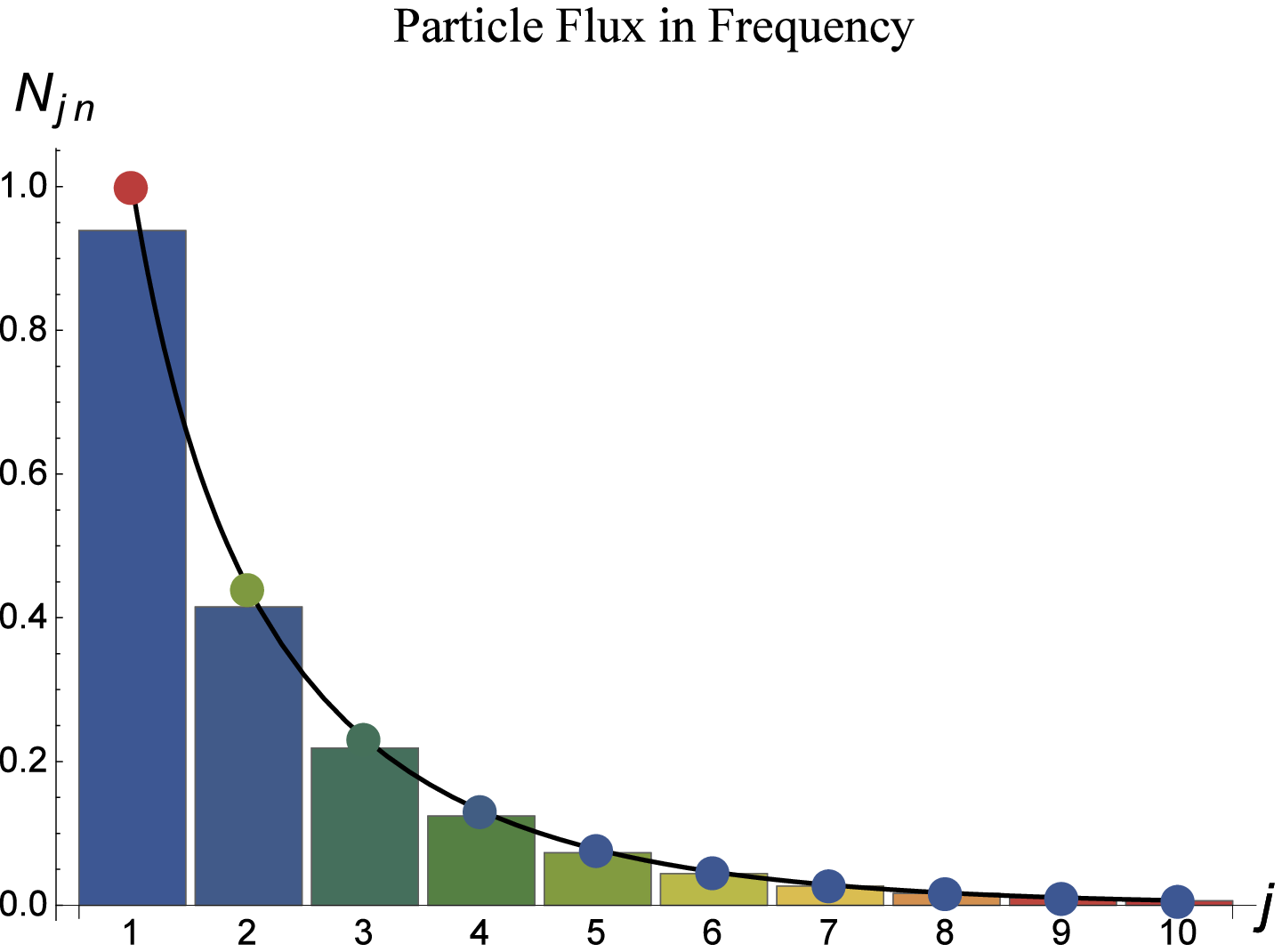}
\caption{\label{fig:pfFREQ}Black mirror particle flux in frequency.  The black curve at the top of the plot is
the Planckian distribution as a function of the frequency, Eq.~\ref{Njn-thermal}. Note that the packet $(j=n=1)$ is in both Figs.  Here $n=1$, $\epsilon = T \ln \phi$, $\kp=1$. }
\end{minipage}
\end{figure}

The black mirror beta can only be exactly analytically packetized at late times.  This is because the exact Planckian result, Eq.~\ref{Njn-thermal}, holds.  Early-time resolution is only tractable numerically by using wave packets of Eq.~\ref{beta-packet} on Eq.~\ref{beta}. As a check, one numerically confirms the late-time exact analytic Planckian distribution, Eq.~\ref{Njn-thermal}.  The reason for first studying the equilibrium mirror before tackling the black mirror head-on should now be clear: since we know a priori that the black mirror produces late-time Planckian particles, we can anchor the late-time numerical precision and accuracy goals against the exact Planckian distribution to gain confidence for the early-time, non-thermal radiation.  
A plot of the time dependence of the particle production for the frequency band $\epsilon = \epsilon_{gr}$ and $\kappa = j = 1$ is given in
 Fig. (\ref{fig:pfTIME}).  The radiation emission is monotonic, i.e. here there is no black hole birth cry\cite{Good:2015nja}.  Since black mirror radiation is black hole evaporation, the black mirror provides a window into the early-time formation phase.

\section*{Acknowledgments}

MRRG thanks Paul Anderson, Yen Chin Ong and Charles Evans for fruitful conversations. MRRG is supported by the Nazarbayev University Social Policy grant.

\end{document}